# Strong Enhancement of Thermal Properties of Copper Films after Chemical Vapor Deposition of Graphene

**Pradyumna Goli[1], Hao Ning[2], Xuesong Li[2], Ching Yu Lu[2], Konstantin S. Novoselov[3] and Alexander A. Balandin[1]**

[1]Nano-Device Laboratory, Department of Electrical Engineering, Bourns College of Engineering, University of California – Riverside, Riverside, California 92521 USA

[2]Bluestone Global Tech, 169 Myers Corners Road, Wappingers Falls, New York 12590 USA

[3]School of Physics & Astronomy, University of Manchester, Oxford Road, Manchester, M13 9PL, UK

## Abstract

We demonstrated that chemical vapor deposition of graphene on Cu films strongly enhances their thermal diffusivity and thermal conductivity. Deposition of graphene increases the thermal conductivity of 9-μm (25-μm) thick Cu films by up to 24% (16%) near room temperature. Interestingly, the observed improvement of thermal properties of graphene coated Cu films is primarily due to changes in Cu morphology during graphene deposition and associated with it temperature treatment rather than graphene's action as an additional heat conducting channel. Enhancement of thermal properties of metal films via graphene coating may lead to applications in electronic circuits and metallurgy.

Graphene is a one-atom-thick material with unusual and highly promising for applications electrical [1-3], thermal [4-5] and mechanical properties [6]. First obtained by mechanical exfoliation from graphite [1-2], graphene is now efficiently grown by chemical vapor deposition (CVD) on copper (Cu) films [7-9]. It was reported that layered graphene – metal composites have enhanced mechanical strength [10]. However, it is not known how deposition of graphene on Cu films affects their thermal properties. Here we demonstrate that CVD of graphene enhances the thermal diffusivity, $\alpha$, and thermal conductivity, $K$, of graphene coated Cu films. Deposition of graphene increases $K$ of 9-$\mu$m (25-$\mu$m) thick Cu films by up to 24% (16%) near room temperature (RT). Interestingly, the enhancement of thermal properties of graphene coated Cu films is primarily due to changes in Cu morphology during graphene deposition and associated with it temperature treatment rather than graphene's action as an additional heat conducting channel. Enhancement of thermal properties of metal films via graphene coating may lead to transformative changes in metallurgy and graphene applications in hybrid graphene – Cu interconnects in Si complementary metal-oxide-semiconductor (CMOS) technology.

Graphene is known to have usually high intrinsic thermal conductivity, which can exceed that of bulk graphite limit of $K\approx2000$ W/mK at RT in sufficiently large high-quality samples [4-5]. However, graphene placement on substrates results in degradation of thermal conductivity to ~600 W/mK owing to phonon scattering on the substrate defects and interface [11]. The benefits of using single-layer graphene (SLG) or few-layer graphene (FLG) as heat spreaders for large substrates are not obvious owing to the small thickness of graphene ($h$=0.35 nm) and possible thermal conductivity degradation by extrinsic effects. Even if $K$ is high, the uniform heat flux, $\Phi=K\times A$, through the cross-sectional area $A=hW$ will be small due to small $h$ ($W$ is the width of the graphene layer). In this Letter, we show that coating copper films with CVD graphene does improve the heat spreading ability of Cu. The enhancement of thermal properties of graphene coated Cu films is primarily due to changes in Cu morphology during graphene deposition and associated with it temperature treatment. Specifically, CVD of graphene results in strong enlargement of Cu grain sizes and reduced surface roughness. A typical grain size in Cu films coated with graphene is larger than that in reference Cu films and in Cu films annealed under the same conditions without graphene deposition.

To demonstrate the effect we used a set of Cu films (thickness $H$=9 μm and $H$=25 μm) with SLG and FLG synthesized on both sides via CVD method (Bluestone Global Tech, Ltd.). As references we used (i) Cu films without graphene or any thermal treatment, and (ii) Cu films annealed under the same conditions as the one used during CVD of graphene. Thus, for comparison we had regular Cu, annealed Cu, Cu with CVD SLG and Cu with CVD FLG. Details of sample preparation are provided in *Methods* section. The reference Cu and Cu-graphene samples were subjected to optical microscopy, scanning electron microscopy (SEM) and atomic-force microscopy (AFM) inspection. The number of atomic planes in graphene films on Cu was verified with micro-Raman spectroscopy (Renishaw In Via). Details of our Raman measurement procedures have been reported by some of us elsewhere [12].

The measurements of the thermal diffusivity were carried out using the "laser flash" method (Netzsch LFA). In conventional configuration, the "laser flash" method gives the cross-plane thermal diffusivity, $\alpha$, of the sample [13]. Since we are mostly interested in the in-plane heat spreading properties of graphene coated Cu films, we altered the experiment by using a special sample holder, which send the thermal energy along the sample. In this approach, the location for the light energy input on one side of the sample and location for measuring the temperature increase on the other side of the sample are at different lateral positions. The latter insures that the measured temperature increase of the sample corresponds to the thermal diffusivity in the in-plane direction. The thermal conductivity was determined from the equation $K=\rho\alpha C_p$, where $\rho$ is the mass density of the sample and $C_p$ is the specific heat of the sample measured separately. Details of the measurements are summarized in *Methods* section. Figure 1 presents a schematic of the experiment, an image of a typical sample with the sample holder, and Raman spectra from two different Cu substrates indicating that one has SLG coating while the other has FLG coating.

[**Figure 1**]

Figure 2 presents the average apparent thermal diffusivity and thermal conductivity in reference Cu films, annealed Cu films, Cu films with CVD graphene and Cu films with CVD FLG. The data are presented for two thicknesses of Cu films: $H$=25 μm and $H$=9 μm. The term *apparent* (another common term is *effective*) emphasizes that $\alpha$ and $K$ values are measured for the whole graphene-Cu-graphene sample. The averaging for each type of sample (e.g. Cu film with SLG) was performed for five locations on each film at each temperature. Two films with the same type of samples were tested. In order to simplify the analyses, in Table I, we provided the average RT values of $\alpha$ and $K$ measured for different samples and locations. The ranges for $\alpha$ and $K$ values for different locations and samples are given in the brackets. The data scatter for different locations was attributed to the sample non-uniformity and film bending, which were unavoidable for large foils (cm scale lateral dimensions) with small thicknesses.

[**Figure 2**]

The obtained $\alpha$ and $K$ of Cu films and their weak temperature dependence are consistent with literature values for bulk Cu, which varies from 385 W/mK to 400 W/mK [14-15]. Electrons are the main heat carriers in Cu while phonons make the dominant contribution in graphene. The strong reduction of $K$ of Cu due to electron scattering from the film top and bottom boundaries is only expected in very thin films where the electron mean-free path (MFP) becomes comparable with $H$ [16]. However, it is known that the grain size in Cu decreases with the decreasing film thickness [15]. For this reason, the size effects can reveal themselves even in relatively thick Cu films with $H \leq 10$ μm [15]. The lower $\alpha$ and $K$ for 9 μm films than those for 25 μm films measured in our experiments are likely related to the grain size effects. The rolling fabrication of Cu films of different thickness (9 μm vs. 25 μm) is also expected to result in variations in the defect densities, grain elongation and orientation, thus, affecting $\alpha$ and $K$.

The most important and unexpected observation from Figure 2 is that $\alpha$ and $K$ are strongly increased in Cu films with graphene or FLG coating compared to reference Cu films or annealed

Cu films. Deposition of graphene results in stronger increase of $\alpha$ and $K$ than annealing under the same conditions. In terms of thermal conductivity, the effect of graphene deposition is particularly pronounced for thinner Cu films ($H$=9 μm). The deposition of SLG on 9-μm Cu film results in about ~22% enhancement of the apparent thermal conductivity as compared to ~12% increase in the annealed samples without graphene. The average enhancement of $K$ and $\alpha$ after deposition of SLG on 25-μm films is less pronounced than that for 9-μm films but still notably larger than for the annealed reference samples. The increase in $\alpha$ and $K$ is not proportional because the thermal treatment during CVD or annealing affects the specific heat as well. It is known that thermal treatment of metals and alloys can noticeably change $C_p$, particularly in the presence of impurities and defects [17].

The overall enhancement of heat conduction properties is very strong and may appear puzzling. The thickness of graphene $h$=0.35 nm is negligibly small compared to $H$=25 μm. For this reason, the thermal resistance $R_{\theta}=L/(KhW)$ of the additional heat conduction channel via graphene will be much larger than via Cu film (here $L$ is the length of the path). Thus, the high thermal conductivity of graphene [5] should not play a significant role in heat spreading ability of Cu foils over large distances ($L$~5 mm) if one considers conventional heat transfer by phonons. The observed enhancement of the apparent $\alpha$ and $K$ can be understood if the thermal data is correlated with the microscopy data presented in Figure 3.

One can see that CVD of graphene results in substantially stronger enlargement of Cu grains than annealing under the same conditions. The graphene CVD and annealing temperature 1030 °C is sufficiently larger than Cu recrystallization temperature of ~227 °C [18]. As a result, annealing accompanied by re-crystallization increases the grain sizes in Cu films, reduces the defect density and improves their mechanical properties [18-19]. Our results indicate that CVD of graphene enhances the Cu grain growth, as compared to regular annealing, by changing the thermal balance during the deposition. Graphene also stops copper evaporation from the surface when the sample is heated during CVD. These conclusions are supported by earlier observations that the substrates and underlays affect the annealing process of Cu and the resulting Cu morphology [20]. It is also in agreement with the grain size data in Cu with CVD graphene and annealed Cu presented in Ref. [21]. Additionally, our SEM studies indicate that CVD of graphene results in ~20% reduction in surface roughness as compared to reference Cu.

**[Figure 3]**

In order to further rationalize the experimental results we estimated the ratio of the average grain sizes, $\tilde{D}/D$, which would provide the relative change in the thermal conductivity, $\Delta K/K$, close to the one observed in the experiments ($\tilde{D}$ is typical grain size in reference Cu film and $D$ is the grain size after CVD of graphene). The electron MFP for thermal transport is $\Lambda$=40 nm at RT [17]. Since $\Lambda << H$, it is reasonable to assume that $K$ is mostly limited by the grain boundary scattering. In this case, one can express the thermal conductivity, $K$, of a polycrystalline metal through that of a single-crystal bulk metal, $K_B$, as [22-24] $K = (1 + \Lambda/D)^{-1} K_B$. Applying this equation to polycrystalline Cu before and after CVD of graphene we derived the following relation

$$\frac{\tilde{D}}{D} = \frac{1 - (\Delta K/K)}{1 + (\Delta K/K)(D/\Lambda)}.$$
(1)

If one assumes that the average grain diameters are in the range $D \approx$1-10 μm, the experimentally measured $\Delta K/K$=0.2 can be achieved for if $\tilde{D}/D$ varies from ~0.13 to 0.016, which corresponds to the grains in reference Cu on the order of $130 - 160$ nm. Since our samples have large variation of the grain sizes it is difficult to assign the mean value. The considered range and change in the diameter by ×10-×100 after CVD is consistent with the microscopy data (see examples in Figure 3 and *Supplementary Information*). It is known that annealing of Cu under different conditions can change the grain size by many orders of magnitude from ~30 nm to 100 mm [19]. Our analysis suggests that the grain size increase can result in the observed enhancement of the thermal conductivity. Variations in the defect densities, e.g. dislocation lines, and grain boundary thickness after CVD of graphene may also affect the $\Delta K/K$.

Although it is clear that the observed strong enhancement of thermal properties of Cu films after CVD of graphene is mostly related to the effect produced by graphene on Cu grains one cannot completely exclude other possible mechanisms of heat conduction, which might be facilitated by graphene. It has been recently suggested theoretically that plasmons and plasmon-polaritons can strongly enhance the heat transfer in graphene and graphene-covered substrates [25-26]. The fact in our measurements the samples are heated by the light flash with the wide spectrum leaves this possibility open. The plasmon contribution would come in addition to the phonon heat conduction in graphene.

Practical significance of our results can be illustrated by the following considerations. Carbon additives have long been used in metallurgy, e.g. in steel smelting, as alloying elements distributed through the volume. Carbon alloying allows one to vary the hardness and strength of the metal [18]. Our results show that CVD of one-atom-thick graphene layer on the surface of metal foils can have a pronounced effect on its thermal properties. This is a conceptually different approach for carbon use in metallurgy. In another application domain, Cu became crucially important material for interconnects in Si CMOS technology by replacing Al. Main challenges with continuous downscaling of Si CMOS technology include electromigration in Cu interconnects and heat dissipation in the interconnect hierarchies separated from a heat sink by many layers of dielectrics [27]. It has been demonstrated that the breakdown current density in prototype graphene interconnects exceeds that in metals by $\times 10^3$ [28] and that graphene capping of Cu interconnects increases the current density and reduces resistance [29]. Intersecting hybrid graphene – Cu interconnects have also been proposed [30]. Our present findings add validity to the proposals of the graphene capped Cu interconnects by demonstrating improvement in their heat spreading ability. Taking into account that the next technology nodes will require Cu interconnects with the nm-range thickness [27] one can expect that the effects will be even more pronounced than in the examined μm-range thickness films.

**METHODS**

**Sample Preparation:** The purity of 25-μm thick copper is 99.9 % and that of 9-μm thick copper is above 99.99 %. Graphene is synthesized in a low-pressure CVD system following the method described in Refs. [7-8]. A copper substrate is heated up to 1030 °C under hydrogen and then methane is introduced for graphene growth. The samples with SLG and FLG are synthesized by controlling the cooling rate. For the case of SLG, the copper substrate is cooled from 1030 °C to RT within 20 minutes while for FLG the cooling time is about 10 hours. The annealing of copper for reference samples is performed with the same heating and cooling process as that of SLG synthesis but no methane addition during the process.

**Measurement Details:** The "laser flash" technique (LFT) is a transient method that directly measures $\alpha$. The specific heat, $C_p$, is measured independently with the same instrument using Cu reference. To perform LFT measurement, each sample was placed into a special stage and sample holder (see Figure 1) that fitted its size. The bottom of the stage was illuminated by a flash of a xenon lamp (wavelength $\lambda$=150 – 2000 nm) with the energy pulse of 1 J for 0.3 ms. The temperature of the opposite surface of the sample was monitored with a cryogenically cooled InSb IR detector. The design of the "in-plane" sample holder ensured that heat traveled ~5 mm inside Cu film along its plane, which is a much larger distance than its 25 μm thickness, and thus, ensuring the in-plane values for $\alpha$ and $K$. The temperature rise as a function of time, $\Delta T(t)$, was used to extract $\alpha$. The specific heat, $C_p$, was measured with LFT by comparing $\Delta T(t)$ of the sample to that of a reference sample under the same experimental conditions ($\alpha$ of the reference Cu was ~0.39 J/gm×K at RT). Annealing or CVD of SLG increased $C_p$. The increase of specific heat with CVD of graphene or FLG was attributed to morphological changes induced by high temperature during the CVD and the fact that specific heat of graphite, $C_p$=0.71 J/gm×K, is larger than that of Cu. The accuracy of LFT measurement with Netzsch instruments is ~1-3%. The thermal conductivity was determined from the equation $K=\rho\alpha C_p$, where $\rho$ is the mass density of the sample.

*Acknowledgements*

The work at UC Riverside was supported, in part, by the National Science Foundation (NSF) project ECCS-1307671 on engineering thermal properties of graphene, by DARPA Defense Microelectronics Activity (DMEA) under agreement number H94003-10-2-1003, and by STARnet Center for Function Accelerated nanoMaterial Engineering (FAME) – Semiconductor Research Corporation (SRC) program sponsored by MARCO and DARPA.

## Author Contributions

A.A.B. led the thermal data analysis and wrote the manuscript; K.S.N. coordinated the project, contributed to data analysis and manuscript preparation; H.N, X.L. and C.Y.L. prepared the samples; P.G. performed material characterization, thermal measurements and contributed to data analysis.

## Author Information

The authors declare no competing financial interests. Correspondence and requests for materials should be addressed to (A.A.B.) Alexander.Balandin@ucr.edu and (K.S.N) Konstantin.Novoselov@manchester.ac.uk

**FIGURE CAPTIONS**

**Figure 1: Samples and the measurement setup.** (a) Schematic of the modified "laser flash" experimental setup for measuring in-plane thermal diffusivity. (b) Cu film coated with CVD graphene placed on the sample holder. (c) Back side of the sample holder with the slits for measuring temperature. Cu film is seen through the openings. (d) Raman spectrum of graphene and few-layer graphene on Cu. The data is presented after background subtraction.

**Figure 2: Thermal diffusivity and thermal conductivity of graphene coated copper films.** Thermal diffusivity of reference Cu film, annealed Cu, Cu with CVD graphene, and Cu with CVD FLG (top panels). Thermal conductivity of reference Cu film, annealed Cu, Cu with CVD graphene, and Cu with CVD FLG (bottom panels). The data are shown for Cu films with $H$=9 μm and $H$=25 μm. Note that CVD of graphene and FLG results in stronger increase in the apparent thermal conductivity of graphene-Cu-graphene samples than annealing of Cu under the same conditions.

**Figure 3: Optical and scanning electron microscopy of Cu and graphene coated Cu.** Optical image of the surface of Cu film (a); annealed Cu film (b); and Cu film with CVD graphene (c). SEM image of the surface of Cu film (d); annealed Cu film (e); and Cu film with CVD graphene (f). Note that deposition of graphene substantially increases the Cu grain size.

**Table I: Thermal Diffusivity and Thermal Conductivity of Graphene Coated Cu Films**

| Cu samples | 9 μm | 9 μm annealed | 9 μm with SLG | 9 μm with FLG | 25 μm | 25 μm annealed | 25 μm with SLG | 25 μm with FLG |
|---|---|---|---|---|---|---|---|---|
| $\alpha$ (mm$^2$S$^{-1}$) | 84 | 90.7 (87 − 93) | 89.6 (88 − 93) | 95.5 (91 − 99) | 90 | 91.2 (91 − 92) | 97.6 (95 − 100) | 98.4 (98 − 99) |
| $K$ (W/mK) | 290 | 329.5 (319 − 340) | 369.5 (361 − 379) | 364.3 (346 − 378) | 313 | 337.2 (320 − 358) | 363.0 (354 − 374) | 376.4 (372 − 377) |
| $\Delta\alpha/\alpha$ (%) | − | 7.4 | 6.3 | 12.0 | − | 1.3 | 7.8 | 8.5 |
| $\Delta K/K$ (%) | − | 11.9 | 21.5 | 20.4 | − | 7.2 | 13.8 | 16.9 |

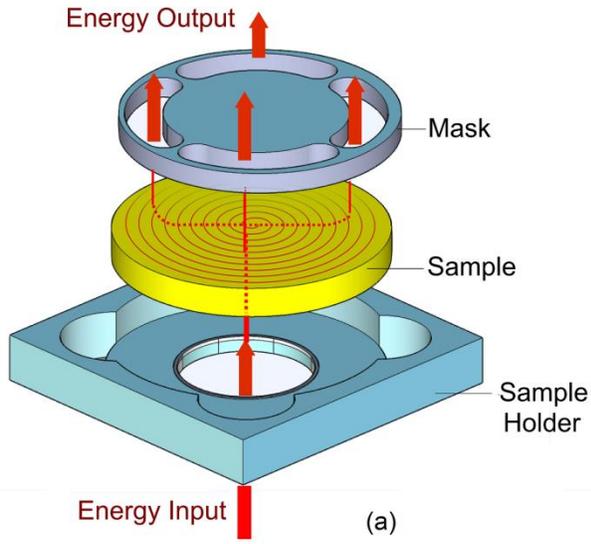

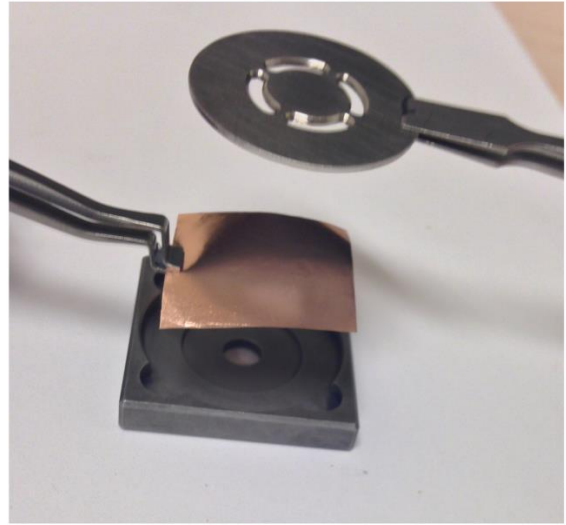

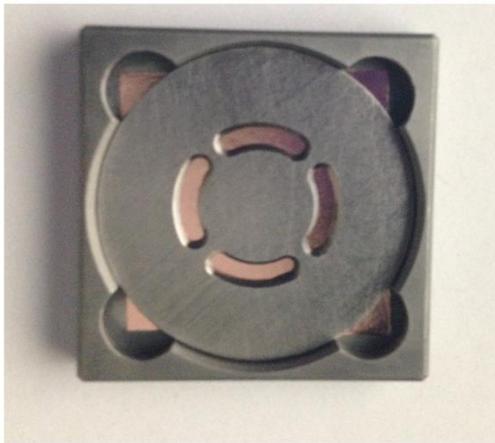

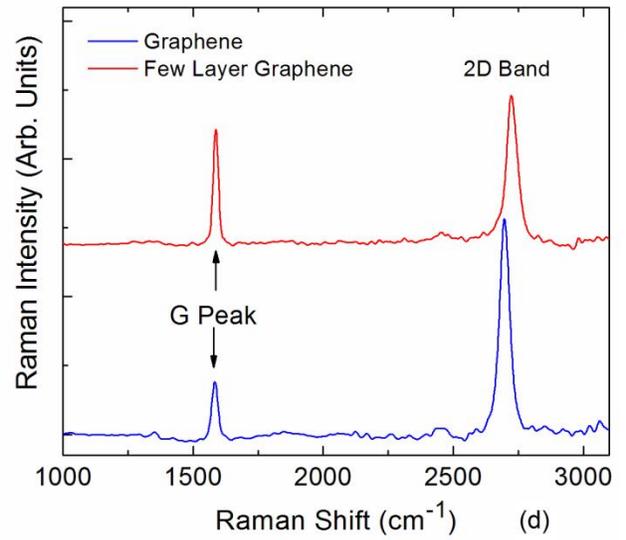

Figure 1

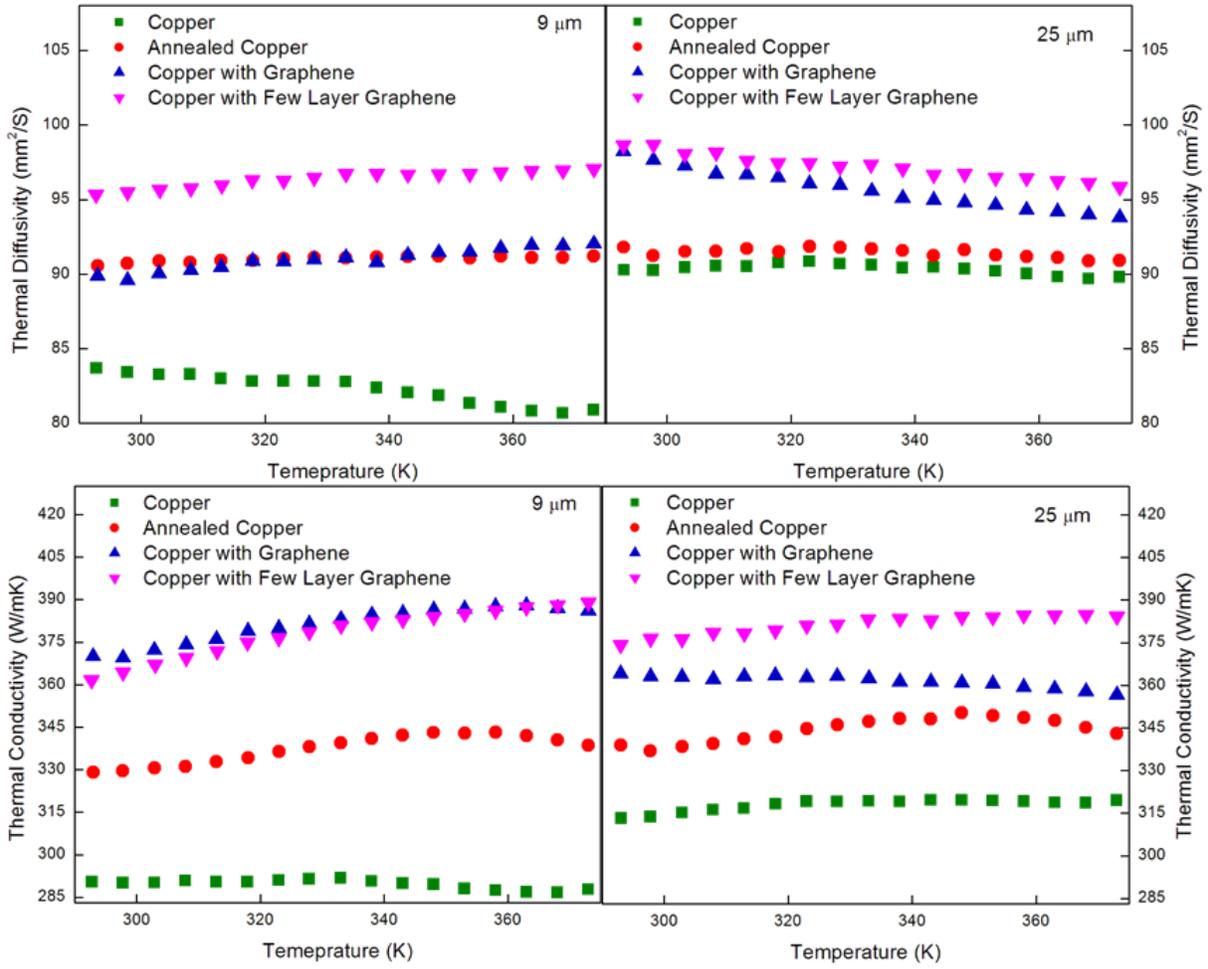

Figure 2

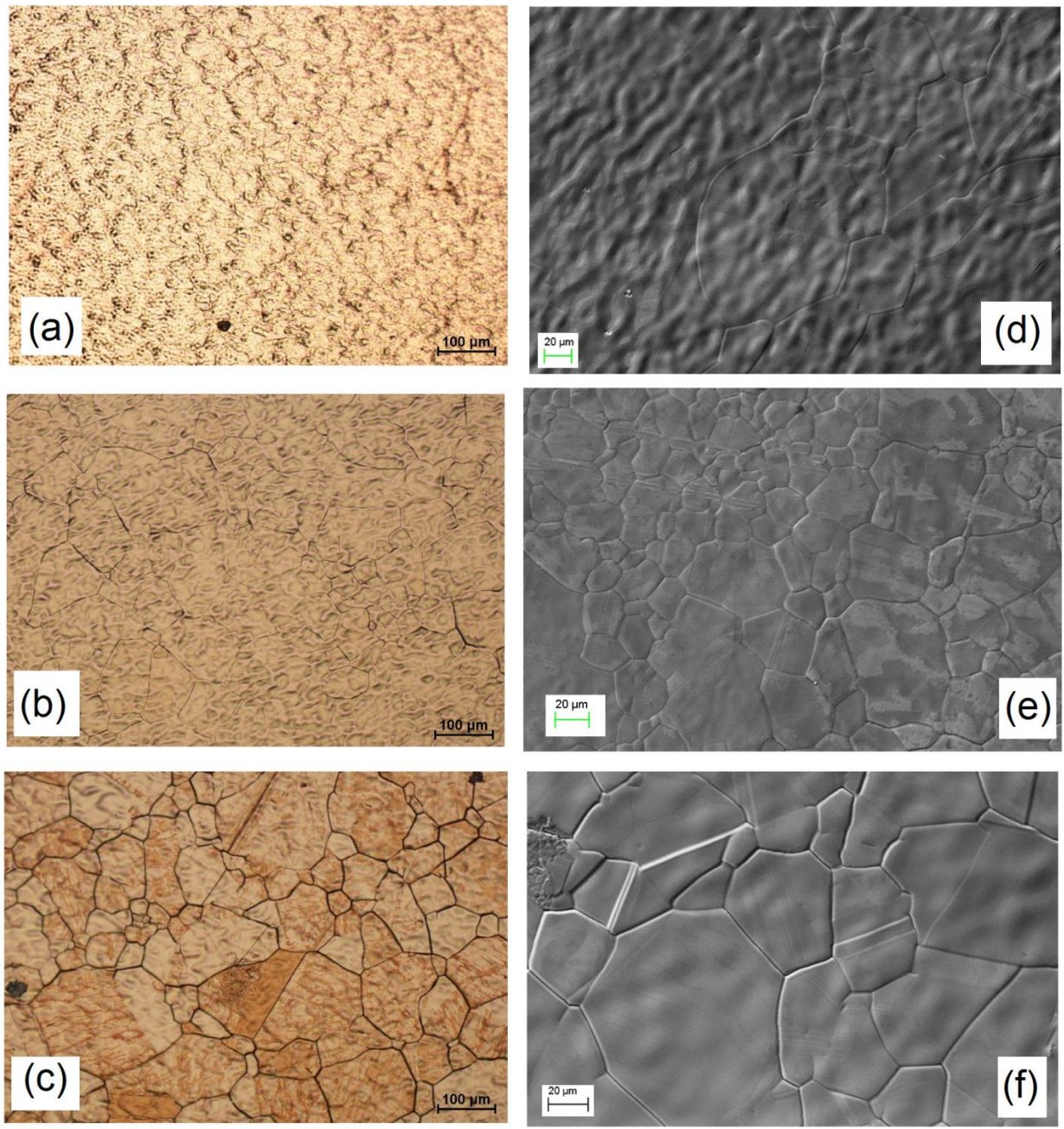

Figure 3